\documentclass[final,3p,times,twocolumn]{elsarticle}

\usepackage{amssymb}
\usepackage{siunitx}
\usepackage{caption}
\usepackage{subcaption}

\journal{Nuclear Instruments and Methods in Physics Research A}

\begin{document}

\begin{frontmatter}



\title{Characterisation and Simulation of stitched CMOS Strip Sensors}
\author{Naomi Davis\corref{cor1}\fnref{DESY}}
\ead{naomi.davis@desy.de}
\cortext[cor1]{Corresponding author.}
\affiliation[DESY]{organization={Deutsches Elektronen Synchrotron DESY}, 
            addressline={Notkestr. 85}, 
            postcode={22607 Hamburg}, 
            country={Germany}}
\author[DESY]{Jan-Hendrik Arling}
\author[TU_Dortmund]{Marta Baselga}
\author[Freiburg,CERN]{Leena Diehl}
\author[Bonn]{Jochen Dingfelder}
\author[DESY]{Ingrid-Maria Gregor}
\author[Freiburg]{Marc Hauser}
\author[Bonn]{Fabian Hügging}
\author[Bonn,Dectris]{Tomasz Hemperek}
\author[Freiburg]{Karl Jakobs}
\author[FH_Dortmund]{Michael Karagounis}
\author[Freiburg]{Roland Koppenhöfer}
\author[TU_Dortmund]{Kevin Kröninger}
\author[Freiburg]{Fabian Lex}
\author[Freiburg]{Ulrich Parzefall}
\author[Freiburg,Littlefuse]{Arturo Rodriguez}
\author[TU_Dortmund]{Birkan Sari}
\author[Freiburg,CERN]{Niels Sorgenfrei}
\author[DESY]{Simon Spannagel}
\author[Freiburg]{Dennis Sperlich}
\author[Bonn]{Tianyang Wang}
\author[TU_Dortmund]{Jens Weingarten}
\author[Freiburg]{Iveta Zatocilova}


\affiliation[Freiburg]{organization={Physikalisches Institut, University of Freiburg},
            addressline={Hermann-Herder-Straße 3},
            postcode={79104 Freiburg},
            country={Germany}}

\affiliation[Bonn]{organization={Physikalisches Institut, University of Bonn},
            addressline={Nussallee 12}, 
            postcode={53115 Bonn},
            country={Germany}}

\affiliation[TU_Dortmund]{organization={Physik E4, TU Dortmund},
            addressline={Otto-Hahn-Strasse 4a},
            postcode={44227 Dortmund},
            country={Germany}}

\affiliation[FH_Dortmund]{organization={Fachhochschule Dortmund},
            addressline={Sonnenstraße 96},
            postcode={44139 Dortmund},
            country={Germany}}

 \affiliation[CERN]{organization={CERN},
        addressline={Esplanade des Particules 1},
        postcode={1211 Meyrin},
        country={Switzerland}}       

 \affiliation[Littlefuse]{organization={Littlefuse},
        addressline={Edisonstraße 15},
        postcode={68623 Lampertheim},
        country={Germany}}       

 \affiliation[Dectris]{organization={DECTRIS AG},
        addressline={Täfernweg 1},
        postcode={5405 Baden},
        country={Switzerland}}

\begin{abstract}
\noindent In high-energy physics, there is a need to investigate alternative silicon sensor concepts that offer cost-efficient, large-area coverage. Sensors based on CMOS imaging technology present such a silicon sensor concept for tracking detectors.

\noindent The CMOS Strips project investigates passive CMOS strip sensors fabricated by LFoundry in a \SI{150}{\nm} technology.
By employing the technique of stitching, two different strip sensor formats have been realised. 
The sensor performance 
is characterised based on measurements at the DESY II Test Beam Facility. 
The sensor response was simulated utilising Monte Carlo methods and electric fields provided by TCAD device simulations. 

\noindent This study shows that employing the stitching technique does not affect the hit detection efficiency. 
A first look at the electric field within the sensor and its impact on generated charge carriers is being discussed. 
\end{abstract}




\end{frontmatter}


\section{Introduction}
\label{sec:introduction}

Particle tracking detectors in high-energy physics primarily rely on silicon sensors as tracking devices.
With the active area of tracking detectors constantly increasing, the high-energy physics community faces a challenge in scaling production volumes and managing costs to cover large-area detectors with high-resolution, radiation-hard silicon sensors. 
Based on this challenge, sensors for particle tracking are only produced by a few foundries, creating the risk of a single-vendor scenario.

This study aims to tackle this challenge by exploring strip sensors based on CMOS imaging technology. As CMOS imaging technology is widely used for industrial and commercial silicon sensor production, it can offer access to fast- and large-scale production by various vendors. In addition, one can profit from the larger wafer size available in commercial processes. 

On the other hand, the maximum reticle size in the order of $\mathcal{O}(\SI{1}{\cm\squared})$ in industrial sensor production poses a remaining challenge for fabricating different sensor formats to achieve large-area coverage in tracking detectors. To circumvent this issue, a sensor is divided into individual parts matching the maximum reticle size, which are illuminated separately on the wafer with a slight overlap. This technique is referred to as \textit{stitching}~\cite{stitching} in CMOS circuit production.  

\section{CMOS Strip Sensors}

The CMOS Strips project is investigating stitched, passive CMOS strip sensors fabricated by LFoundry~\cite{LFoundry} in a \SI{150}{nm} technology.
The sensors are produced on a p-type, \num{8}-inch, Float-Zone wafer with a resistivity between \SI{3}{\kohm} and \SI{5}{\kohm}. 
In addition, the wafer underwent an additional production step at IZM Berlin, where it was thinned down to a thickness of \SI{150(10)}{\um} from the backside and a laser-annealed p$^+$ layer with a closing metallisation layer was implemented as backside implant. 
The stitching technique enables the realisation of a long (\SI{4.1}{\cm}) and a short (\SI{2.1}{\cm}) strip sensor format.

\begin{figure}[tbp]
     \centering
 \begin{subfigure}[t]{0.7\linewidth}
         \centering
         \includegraphics[width=1.0\linewidth]{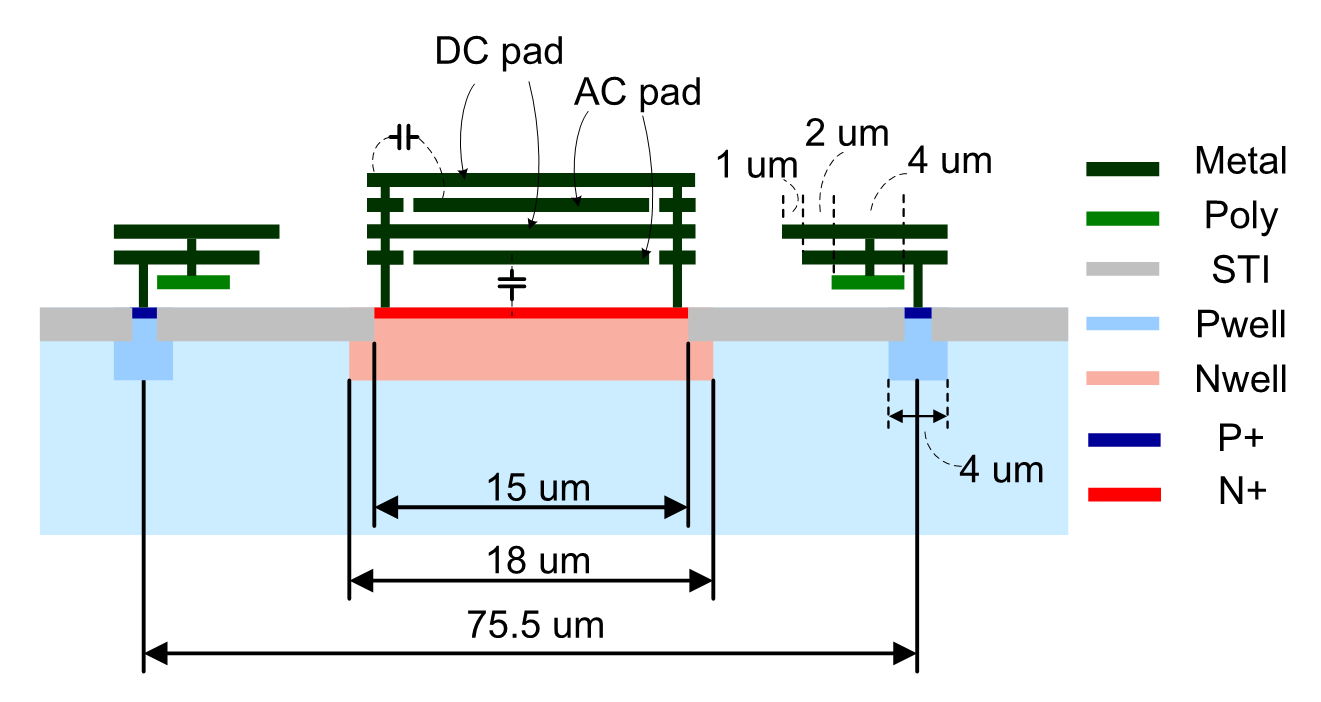}
         \caption{Regular design}
         \label{fig:regular_layout}
     \end{subfigure}
     \begin{subfigure}[t]{0.7\linewidth}
         \centering
         \includegraphics[width=1.2\linewidth]{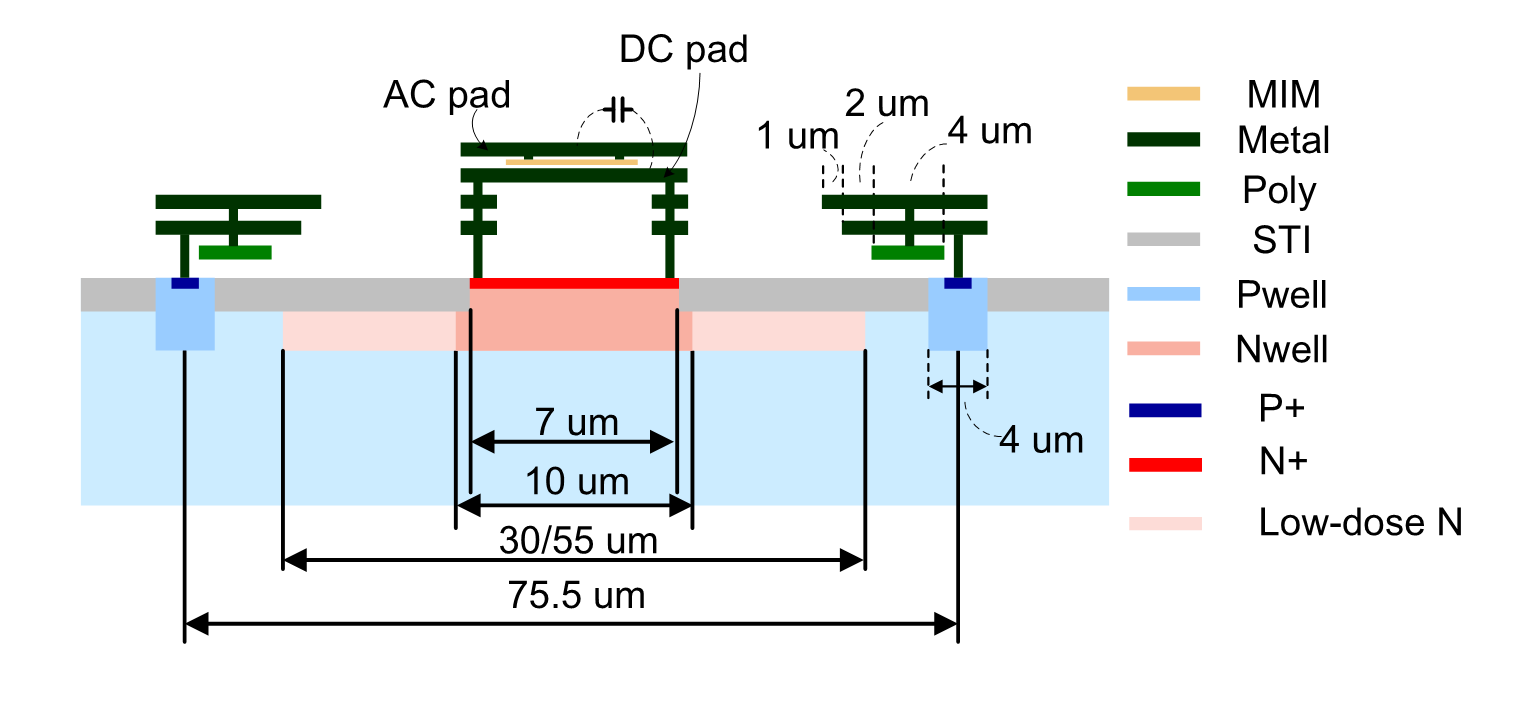}
         \caption{Low Dose design}
         \label{fig:lowdose_layout}
     \end{subfigure}
     \caption{Schematic layout of the Regular (top) and Low Dose \num{30}/\num{55} (bottom) strip implant designs~\cite{DIEHL_Characterization}.}
     \label{fig:layout}
\end{figure}

\begin{figure}[tbp]
    \begin{center}
    \includegraphics[width=0.7\columnwidth]{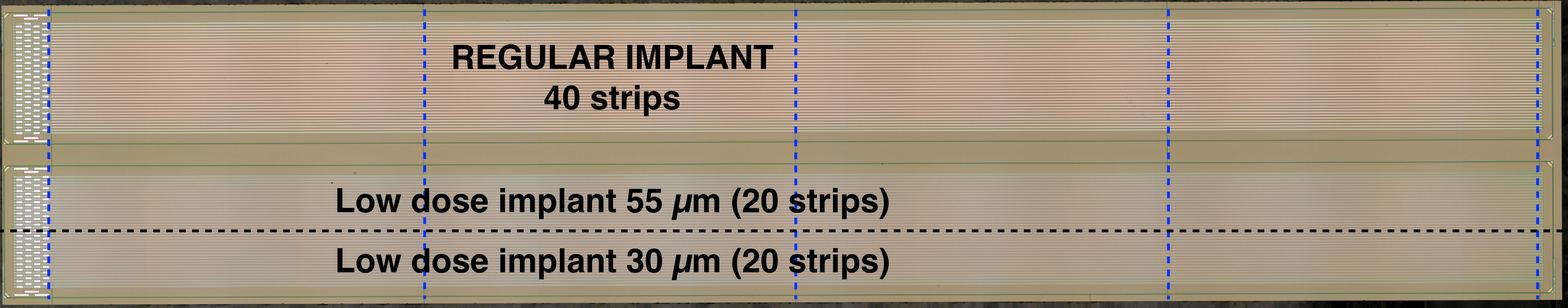}
    \caption{Microscopic picture of a long (\SI{4.1}{\cm}), passive CMOS strip sensor with the different strip implant designs denoted. The blue dashed lines indicate stitching lines across the sensor~\cite{DIEHL_Characterization}.}
    \label{fig:long_strip}
    \end{center}
\end{figure}

The strip implant design varies in doping concentration and width to study various electric field configurations. Figure~\ref{fig:layout} depicts the Regular and the two Low Dose design layouts.
For the Regular design, the strip implant consists of a highly doped n$^+$ layer with a lesser doped n-implant (Nwell) below and has a width of \SI{18}{\um}. The Low Dose designs have a reduced width of \SI{10}{\um} for both the highly doped n$^+$ layer and the lesser doped n-implant. Additionally, these designs are extended by low-doped n-implants (Low-dose N) on each side that reach a total width of \SI{30}{\um} and \SI{55}{\um}, respectively. P-stop implants on either side electrically isolate the strip implants from each other. 
Each sensor comprises \num{40} strips of the Regular design and \num{40} strips of the Low Dose designs, split into Low Dose \num{30} and Low Dose \num{55}, as depicted in Figure~\ref{fig:long_strip}, with a \SI{75.5}{\um} pitch. Figure~\ref{fig:long_strip} also shows where the stitching lines occur along the sensor length, indicated by blue dashed lines. 

In order to evaluate the performance of the different sensor layouts, the sensors are characterised in a test beam environment concerning their hit detection efficiency. In addition, a first simulation of the charge carrier propagation within the sensor is discussed. 
 
\section{Experimental Methods}
\label{sec:methods}
\subsection{Test Beam Measurements}
\label{sec:tb}

The present study is based on data acquired in test beam campaigns at the DESY II Test Beam Facility~\cite{Diener_2019}. 
The facility provides an electron beam with an user-adjustable electron energy from \SI{1}{\GeV} to \SI{6}{\GeV} and offers the ADENIUM~\cite{Liu_2023} beam telescope as a reference for reconstructing particle trajectories through the Device Under Test (DUT). The ADENIUM telescope comprises six ALPIDE~\cite{AGLIERIRINELLA2017583} pixel sensor planes arranged perpendicularly. The strip sensor under investigation is placed in the middle of the telescope with three telescope planes upstream and downstream of the electron beam. Two scintillators, in coincidence, provide trigger signals.

In this study, two different test beam campaigns are being presented. 
The first test beam campaign was conducted in May 2022 with a \SI{3.4}{\GeV} electron beam directed onto an unirradiated strip sensor of the short format. The second campaign in March 2023 featured a \SI{4.2}{\GeV} electron beam directed onto a neutron-irradiated strip sensor of the long format. The irradiation was performed with \SI{23}{\MeV} reactor neutrons at Ljubljana to a fluence of \SI{3e14}{n_{\text{eq}}\per\cm\squared}.  
The DUT is placed in a styrofoam box that is flushed with nitrogen and can be cooled to \SI{-45}{\degreeCelsius} with dry ice pellets to reduce the higher leakage current in irradiated samples. Unirradiated samples were measured at Test Beam area temperature, which averaged to \SI{15}{\degreeCelsius}.

Data acquisition of the strip sensors is performed with the ALiBaVa readout system~\cite{Marco-Hernández:1028151}. It consists of a Daughterboard on which the strip sensors are wire-bonded to \num{128}-channel Beetle Readout chips~\cite{Beetle}. In addition, the ALiBaVa readout system comprises a Motherboard for data acquisition and processing. 

The setup described above allows the study of the strip sensor performance based on the reference track information provided by the beam telescope. 
For track reconstruction and test beam data analysis, the Corryvreckan framework~\cite{Dannheim_2021} is used. A dedicated module has been developed to read and visualise data provided by the ALiBaVa readout system. 

The clustering algorithm used in the presented analysis is based on the Signal-to-Noise Ratio (SNR). It takes a cut on the SNR as an input to define the threshold for seed strips.
Based on this cut value, the sensor hit detection efficiency is studied as a function of the threshold.

\subsection{Simulation}
\label{sec:sim}

Initial sensor simulations were conducted to investigate the strip sensors' electrical properties. Synopsys Sentaurus Technology Computer-Aided Design (TCAD)~\cite{Synopsis} was used to simulate the electrical characteristics of the strip sensors, including the electric field along the sensor thickness at a bias voltage of \SI{100}{V}~\cite{Iveta}. In order to evaluate the impact of the electric field on charge carriers generated within the sensor, the electric field is used as an input to the $\text{Allpix}^2$~\cite{Spannagel_2018} simulation. 

$\text{Allpix}^2$ is a modular framework that allows the simulation of the signal generation within silicon sensors. It enables the study of the charge carrier propagation for the various strip sensor layouts.

\section{Results}
\subsection{Total Hit Detection Efficiency}
\label{sec:total_eff}
\begin{figure}[tbp]
    \begin{center}
    \includegraphics[width=0.9\linewidth]{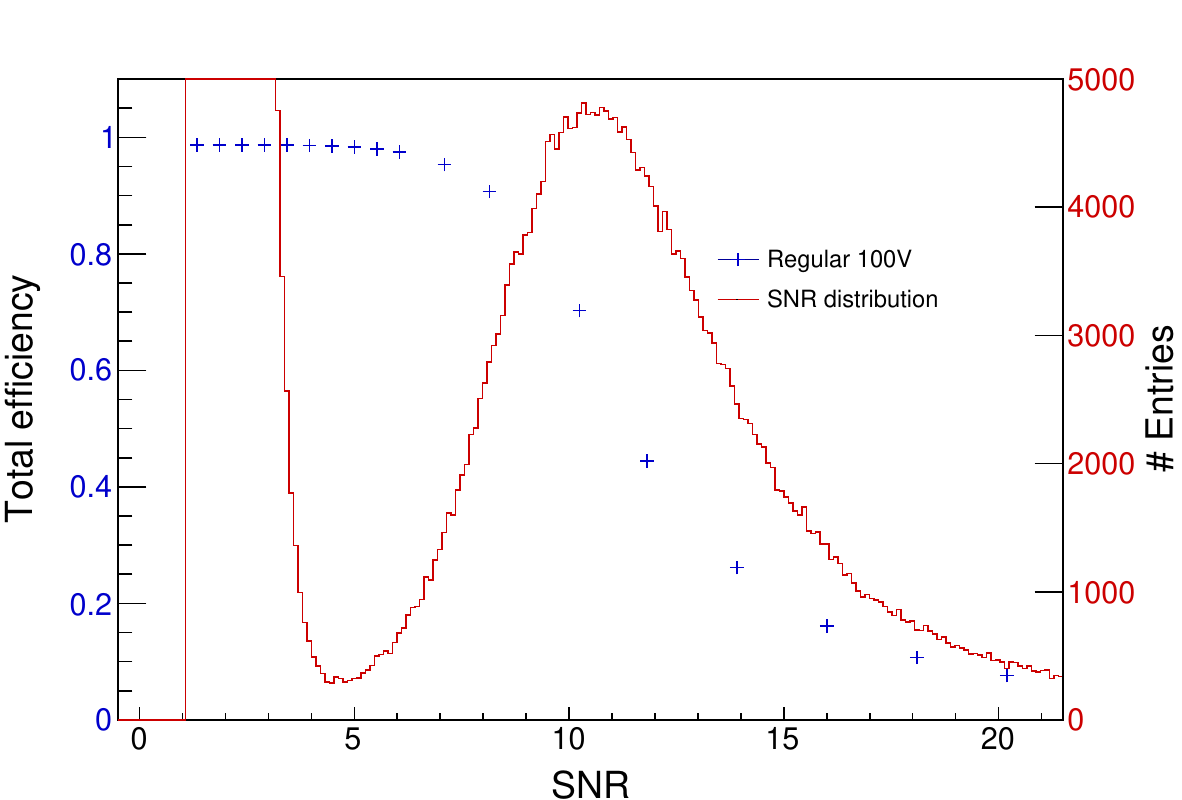}
    \caption{SNR distribution of the Regular design (red, solid), combined with the corresponding hit detection efficiency (blue, crosses) for the unirradiated short strip sensor. The SNR distribution comprises the Gaussian noise distribution and the Landau-shaped signal distribution. The Gaussian noise peak is depicted partially as it is significantly larger than the signal peak.}
    \label{fig:snr}
    \end{center}
\end{figure}

The data acquired in the two test beam campaigns described in section~\ref{sec:methods} is used to evaluate the hit detection efficiency of the measured samples. Additionally, this study investigates whether using the stitching technique affects hit detection efficiency before and after irradiation.
Figure~\ref{fig:snr} shows the hit detection efficiency for the Regular design of the unirradiated sample as a function of the threshold used in the clustering algorithm. The sample is fully depleted at a bias voltage of \SI{100}{V}. The threshold dependency of the total hit detection efficiency for all three strip sensor layouts is published in~\cite{Iveta}. In Figure~\ref{fig:snr}, the threshold dependency is shown with the SNR distribution comprised of a Gaussian-distributed noise and the Landau-shaped signal distribution for the Regular design. 
There is a high-efficiency region for small seed cuts.
The efficiency drops firmly for medium seed cuts while gradually lowering to zero for larger seed cuts, which matches the shape of the SNR distribution. With the seed cut value describing a cut in the SNR distribution, low seed cuts mainly impact the Gaussian distributed noise. As the cut value increases, the Landau-shaped signal distribution begins to be affected. Consequently, reconstructed telescope tracks do not have associated clusters on the DUT, and the hit detection efficiency decreases.

\subsection{In-Strip Hit Detection Efficiency}
\label{sec:in-strip}
\begin{figure}[tbp]
     \centering
 \begin{subfigure}[t]{0.7\linewidth}
         \centering
         \includegraphics[width=\linewidth]{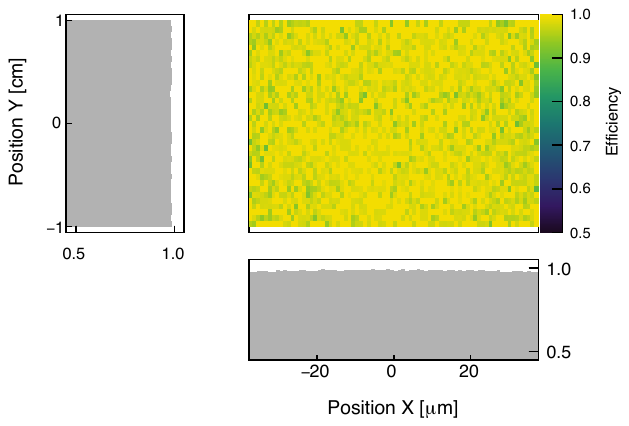}
         \caption{Regular}
     \end{subfigure}
     \begin{subfigure}[t]{0.7\linewidth}
         \centering
         \includegraphics[width=\linewidth]{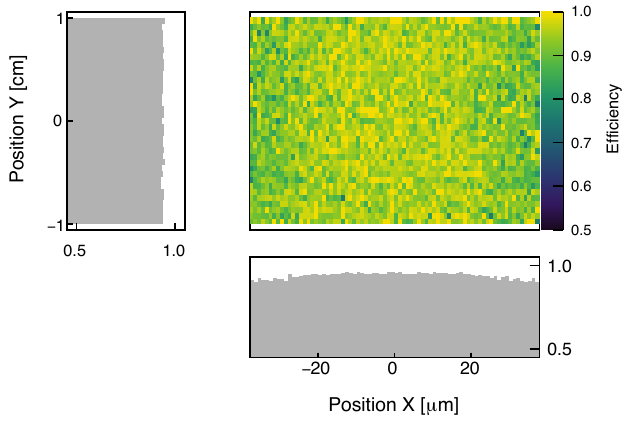}
         \caption{Low Dose 55}
     \end{subfigure}
    \caption{In-strip efficiency of the Regular (top) and the Low Dose 55 (bottom) designs of the unirradiated short strip sensor at a threshold of three. The mean efficiency along the X- and Y-axis is shown. The stitching line across the sensor is located at Y=0.}
    \label{fig:unirradiated}
\end{figure}

Visualising the hit detection efficiency within one strip allows the investigation of any structures along or across the strip, leading to potential inefficiencies.
In order to investigate if the technique of stitching has any impact on the hit detection efficiency, the in-strip efficiency is studied for all three designs. 
The corresponding distribution for the fully depleted Regular and Low Dose 55 design of the unirradiated sample is shown in Figure~\ref{fig:unirradiated}, based on the folded statistics of available strips for the analysis of each design (see~\cite{Iveta} for the results concerning the Low Dose 30 design). 
Overall, the Regular design is more efficient than the Low Dose 55 design. 
The Low Dose 55 design shows a slight decrease in efficiency across the strip towards the strip edge. 
Apart from statistical fluctuations, however, the efficiency is distributed homogeneously along the strip for both designs. 
Therefore, the stitching does not impact the efficiency of the sensor, as the stitching lines highlighted in Figure~\ref{fig:long_strip} do not show in the in-strip efficiency. 

\begin{figure}[tbp]
     \centering
 \begin{subfigure}[t]{0.7\linewidth}
         \centering
         \includegraphics[width=\linewidth]{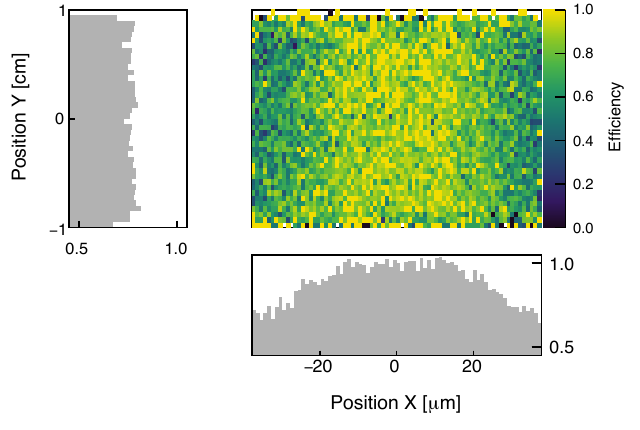}
         \caption{Regular}
     \end{subfigure}
     \begin{subfigure}[t]{0.7\linewidth}
         \centering
         \includegraphics[width=\linewidth]{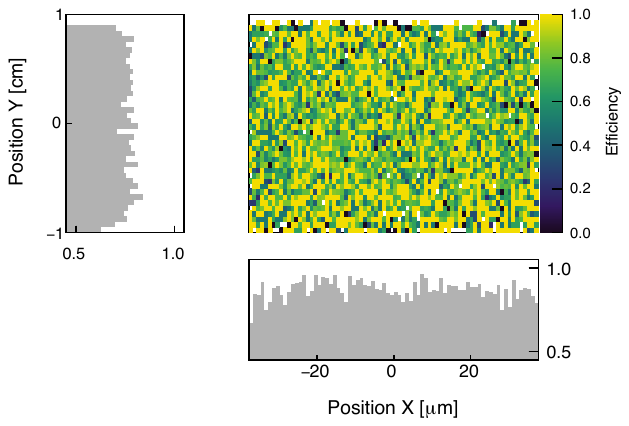}
        \caption{Low Dose 55}
     \end{subfigure}
    \caption{In-strip efficiency of the Regular (top) and the Low Dose 55 (bottom) designs of a neutron-irradiated long strip sensor at a threshold of three. The mean efficiency along the X- and Y-axis is shown. The stitching line across the sensor is located at Y=0.}
    \label{fig:irradiated}
\end{figure}

Figure~\ref{fig:irradiated} shows the in-strip efficiency distribution for the Regular and Low Dose 55 design of the irradiated sample to evaluate if the same finding holds after irradiation. The irradiated sample is depleted at \SI{250}{V}.
Overall, the efficiency decreases after irradiation for both designs. In particular, the Regular design shows a substantial decline in efficiency towards the inter-strip region after irradiation. 
However, the in-strip efficiency shows no sign of the stitching lines along the sensor. 
In addition, no performance differences regarding the hit detection efficiency have been observed between long and short strip samples for the unirradiated and irradiated cases.

\subsection{Simulation}
\label{sec:linegraphs}

\begin{figure}[tbp]
     \centering
 \begin{subfigure}[t]{0.7\linewidth}
         \centering
         \includegraphics[width=\linewidth]{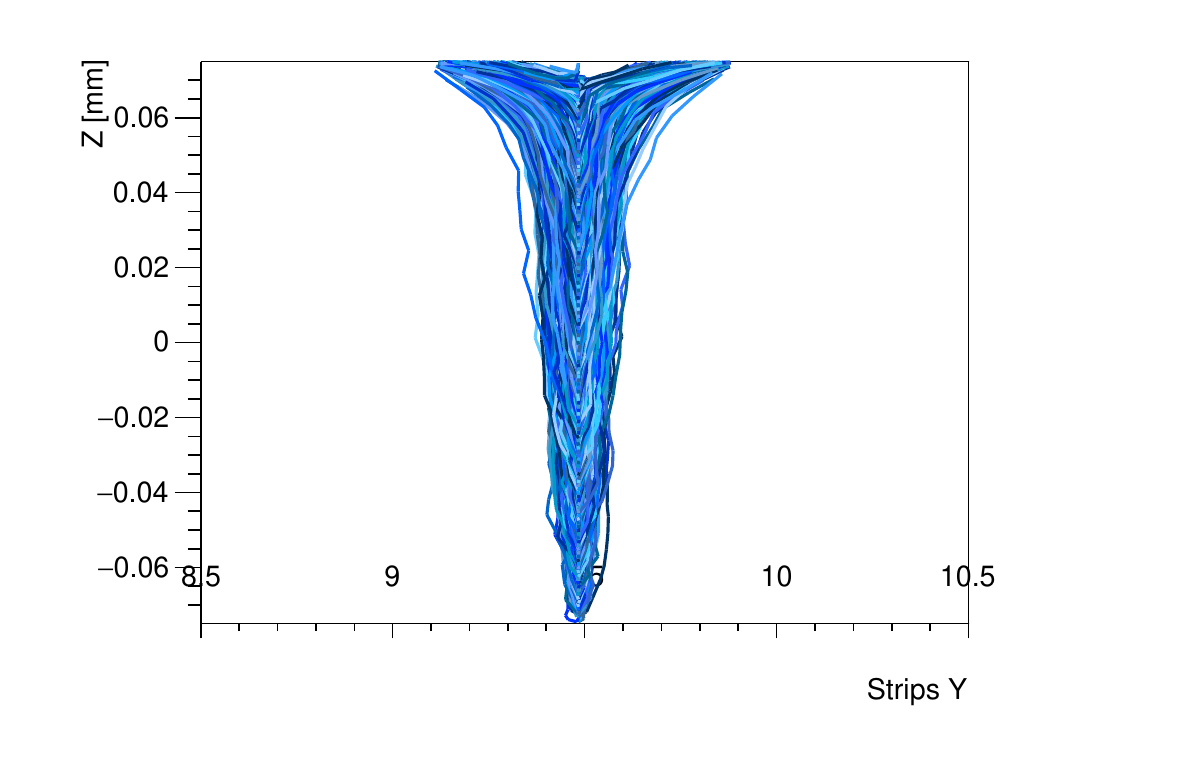}
         \caption{Regular}
     \end{subfigure}
     \begin{subfigure}[t]{0.7\linewidth}
         \centering
         \includegraphics[width=\linewidth]{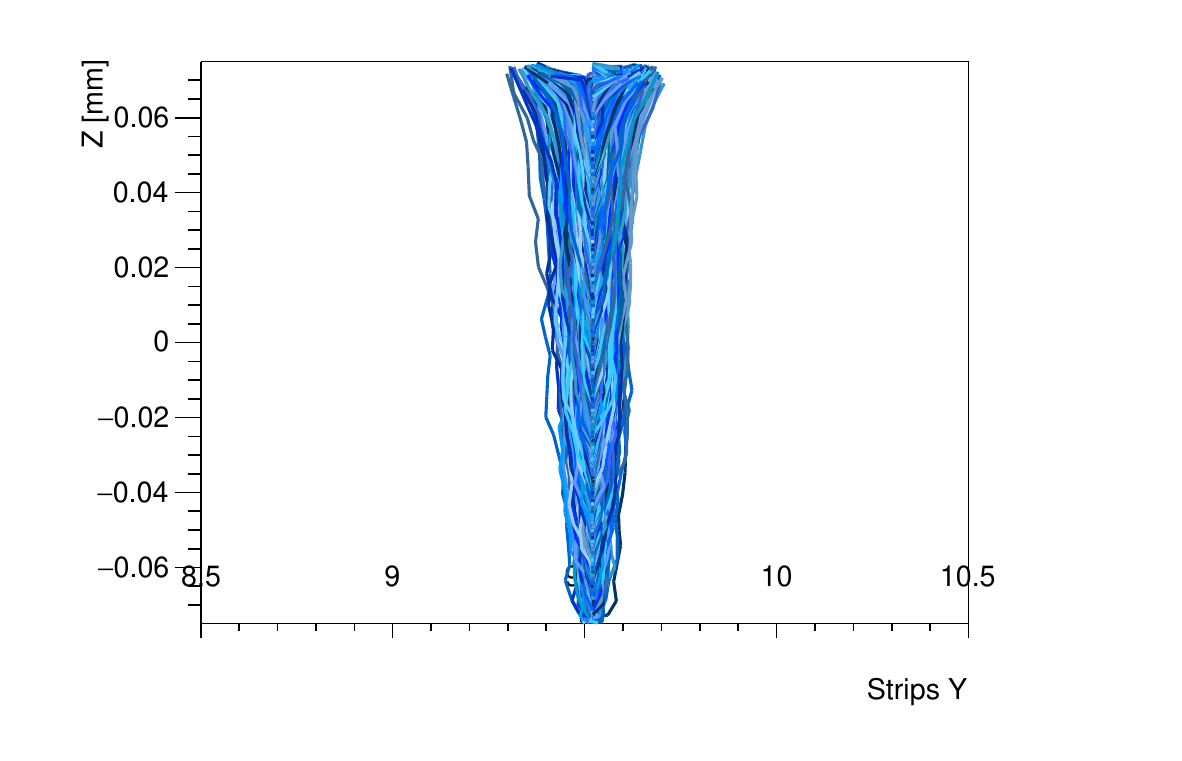}
         \caption{Low Dose 55}
     \end{subfigure}
    \caption{Line graphs visualising the drift path of electrons, created by a traversing MIP, through the sensor thickness to the collection electrode for the Regular (top) and the Low Dose 55 (bottom) designs.}
    \label{fig:linegraph}
\end{figure}

Initial sensor simulations were performed to understand the performance characteristics in the three sensor layouts. The electric field simulated with TCAD is used as an input to the $\text{Allpix}^2$ framework. That allows for visualising the path and collection of the generated charge carriers within the sensor thickness based on the electric field. 
Figure~\ref{fig:linegraph} shows the corresponding line graphs for the Regular and Low Dose 55 designs. They are generated by simulating a Minimum Ionising Particle (MIP) impinging on a strip from the top and ionising the sensor material while travelling through the depleted sensor thickness. The blue lines indicate the generated electrons' drift path towards the collection electrode at the top, located at integer values in $Y$. 
It is visible that there is no substantial difference in the charge carrier propagation between the regular and the Low Dose 55 design. Both designs show significant charge collection by drift (indicated by straight lines). However, the Regular design exhibits a stronger drift towards the collection electrode between the strips than the Low Dose 55 design, due to a larger electric field magnitude near the collection electrode for the Regular design. 
The impact of the different electric field configurations on the various sensor layouts and their final performance needs further investigation. 
 
\section{Conclusion}
\label{sec:conclusion}
Stitched, passive CMOS strip sensors represent an alternative silicon sensor concept that can be used for charged particle tracking. Two samples with three sensor designs have been characterised in a test beam environment. Overall, the Regular implant design shows the best performance regarding hit detection efficiency. The efficiency is lower for the Low Dose designs and declines after neutron-irradiation for all designs. The irradiation of the strip sensors also causes an efficiency decline in the inter-strip region in the Regular design. 

The stitching technique was used in the sensor production to achieve large sensor formats for large-area coverage in tracking detectors. The investigation of the in-strip efficiency has shown that stitching does not impact the hit detection efficiency. The same holds after sensor irradiation. 

Finally, a first simulation of the sensor response was conducted with the $\text{Allpix}^2$ framework. The line graph tool revealed no substantial differences between the Regular and Low Dose 55 strip implant designs. 
Further investigation is required to understand the performance differences between the sensor layouts.
\section{Acknowledgements}
\label{sec:acknowledgements}

The measurements leading to these results have been performed at the Test Beam Facility at DESY Hamburg (Germany), a member of the Helmholtz Association (HGF).



\bibliographystyle{elsarticle-num} 
\bibliography{refs}





\end{document}